\documentclass[aps,pra,showpacs,twocolumn,amsmath,amssymb,superscriptaddress,footinbib]{revtex4}

\ifx\pdfoutput\undefined 
\else 
\fi

\usepackage[english]{babel}
\usepackage{latexsym}

\usepackage{graphics}
\usepackage{subfigure}
\usepackage{epsfig}
\usepackage{amsfonts}
\usepackage{amssymb}
\usepackage{amsmath}
\usepackage{bbm}
\usepackage{color}

\newcommand{\ie}{{\it i.e.}}

\newcommand{\via}{{\it via}~}

\newcommand{\beq}{\begin{equation}}
\newcommand{\eeq}{\end{equation}}

\newcommand{\bea}{\begin{eqnarray}}
\newcommand{\eea}{\end{eqnarray}}

\newcommand{\bra}[1]{\ensuremath{\langle#1|}}

\newcommand{\ket}[1]{\ensuremath{|#1\rangle}}

\newcommand{\ew}[1]{\ensuremath{\langle #1\rangle}}

\newcommand{\eins}{\ensuremath{\mathbbm 1}}

\newcommand{\figref}[1]{Fig.~\ref{#1}}
\newcommand{\LN}{{\rm LN}}

\begin{document}

\title{Quantum State Transfer in Spin-$1$ Chains}

\author{O.~Romero-Isart}
\affiliation{Departament de F{\'\i}sica, Grup F{\'\i}sica
Te\`orica, Universitat Aut\`onoma de Barcelona, E-08193
Bellaterra, Spain.}
\author{K.~Eckert}
\affiliation{Departament de F{\'\i}sica, Grup F{\'\i}sica
Te\`orica, Universitat Aut\`onoma de Barcelona, E-08193
Bellaterra, Spain.}
\author{A.~Sanpera}
\affiliation{Departament de F{\'\i}sica, Grup F{\'\i}sica
Te\`orica, Universitat Aut\`onoma de Barcelona, E-08193
Bellaterra, Spain.}

\affiliation{ICREA: Instituci\'o Catalana de Recerca i Estudis
Avan\c cats.}

\date{\today}

\begin{abstract}
We study the transfer of quantum information through a Heisenberg spin-$1$ chain prepared in its ground state. We measure the efficiency of such a quantum channel {\em via} the fidelity of retrieving an arbitrarily prepared state and {\em via}  the transfer of quantum entanglement. The Heisenberg spin-1 chain has a very rich quantum phase diagram.  We show that the phase boundaries are reflected in sharp variations of the transfer efficiency. In the vicinity of the border between the dimer and the ferromagnetic phase (in the conjectured spin-nematic region), we find strong indications for a qualitative change of the excitation spectrum. Moreover, we identify two regions of the phase diagram which give rise to particularly high transfer efficiency; the channel might be non-classical even for chains of arbitrary length, in contrast to spin-$1/2$ chains.
%

\end{abstract}


\maketitle

\begin{figure}[b]
\begin{center}
\includegraphics[width=0.98\columnwidth]{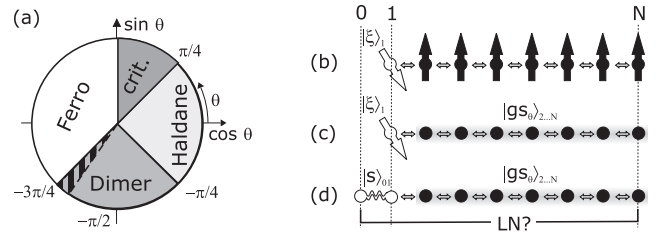}
\end{center}
\caption{ (a) Quantum phases of the spin-$1$ chain. (b) Scheme for
state transfer with an underlying ferromagnetic state, (c) with
the channel initialized to the ground state of $\hat H(\theta)$,
and (d) transfer of quantum entanglement.}
\label{fig:phases_scheme}
\end{figure}

Since the early works of Osborne {\it et al.}~\cite{osborne:2002}
and Osterloh {\it et al.} \cite{osterloh:2005}, displaying an
intertwined relation between entanglement and quantum phase
transitions, the fields of condensed matter and quantum
information have developed a strong synergy, further motivated by
the spectacular advances reached in the area of ultracold atomic
physics and ion traps \cite{jaksch:1998}. Among the broad scope of
problems that nowadays can be addressed with ultracold atomic
gases, spin models are particularly appealing. These relatively
simple models exhibit the most fundamental physics associated with
magnetic ordering, criticality, and quantum phase transitions, but mostly
lack an analytical solution. Though spin
models have been constructed as idealizations or toy
models of real systems, ultracold atoms allow for an almost
perfect realization of many of them.
%
%
An example is the one-dimensional (1D) spin-$1$ system,
which can be realized through confining an $S=1$ spinor condensate
in an optical lattice \cite{demler:2002,yip:2003}. Restricting to
nearest-neighbor interactions, the most general isotropic
Hamiltonian for the spin-$1$ chain is the bilinear-biquadratic
Hamiltonian (BBH) \beq \hat H(\theta)=J\sum_{<ij>} \left[
\cos\theta (\vec S_i \vec S_j)+\sin\theta (\vec S_i\vec S_j)^2
\right]. \label{Heisenberg} \eeq Here $\vec
S_i=(S_i^x,S_i^y,S_i^z)$ are the spin operators on the $i$th site,
and $\cos\theta$ ($\sin\theta$) gives the strength of the bilinear
(biquadratic) coupling. The properties of the ground state as well
as of the excitations are determined by the angle $\theta$. The
phase diagram is shown in Fig.~\ref{fig:phases_scheme}(a).
In the whole range $-3\pi/4<\theta<\pi/2$, the ground state is
antiferromagnetic, \ie, has vanishing magnetization:
$\vec{M}=\langle \sum_i \vec S_i\rangle=0$. Since the Haldane conjecture that 1D isotropic antiferromagnets with integer
spin must have a unique massive, \ie, gapped, ground state with
exponentially decaying correlations, the BBH has been extensively
studied. The Haldane conjecture was rigorously proven for
the AKLT point ($\tan\theta=1/3$), for which Affleck {\it
et al.}~explicitly constructed the ground state and proved the existence of a gap
\cite{aklt:1987}. At $\theta=\pi/4$ (Uimin-Lai-Sutherland point
\cite{uimin:1970,sutherland:1975}) the system enters into a
critical (gapless) phase ($\pi/4\leq\theta<\pi/2$) with unique ground state
and diverging correlation length. At $\theta=-\pi/4$ the gap vanishes \cite{tak:1982}, but
it reopens for $\theta<-\pi/4$, where the systems enters into a
dimerized phase. The exactly solvable point
$\theta=-3\pi/4$ marks the border to the ferromagnetic phase (characterized by $\vec{M}\neq0$). The existence of a small
spin-nematic region between the dimerized phase and
$\theta=-3\pi/4$ is actively discussed since a
conjecture of Chubukov \cite{chubukov:1991}.

In the field of quantum information, spin chains have been
intensively studied regarding their usefulness as quantum channels
\cite{bose:2003:b,subr:2004,christandl:2004}. Attention has been
devoted nearly exclusively to spin-$1/2$ chains, where it has been
shown that a general quantum state can be transfered with
relatively high fidelity between the two endpoints of a
ferromagnetic chain with nearest-neighbor interactions
\cite{bose:2003:b,subr:2004}.
As models with more complex ground states, spin-$1/2$ chains in
the vicinity of a quantum phase transition \cite{hartmann:2006},
spin-$1/2$ ladders \cite{li:2005} and Peierls distorted chains
\cite{huo:2006} have been employed as quantum channels.

Here we investigate the usefulness of the spin-$1$ chain as a
quantum channel. To this aim, we study the transmission of an
arbitrary quantum state, either pure or entangled with a further
subsystem, through the chain prepared in its ground state {\it for
the whole phase diagram}. Compared to previous works studying
either the ground state magnetic order or the excitation spectrum
of the BBH, our approach combines both aspects simultaneously. We
find that the transfer does strongly depend on the nature of the
ground state. When crossing the phase boundaries, the quality of
the transfer changes sharply. Inside the Haldane phase (around
the AKLT point) transfer is very inefficient. On the other hand,
high transfer fidelities can be achieved by preparing the
spin-$1$ chain in either the dimerized (gapped) or the critical
(gapless) phase. In those cases, transfer fidelities are clearly
larger than for a ferromagnetic initial state. Finally, we find a
strong reduction of the transfer efficiency for
$\theta\gtrsim-3\pi/4$, where a spin-nematic phase has been
conjectured. This finding supports recent studies showing a
qualitative change in the low-energy excitations as compared to
the dimerized phase.

Our scheme for quantum communication generalizes the one usually
employed for spin-$1/2$ systems (Fig.~\ref{fig:phases_scheme}(b)):
we consider a chain of $N$ sites with the first spin in an
arbitrary state $\ket{\xi}_1=\sum_{m=0,\pm 1}\xi_m\ket{m}_{1}$
($S_z\ket{m}=m\ket{m}$, $\sum_m|\xi_m|^2=1$) and decoupled from
the rest of the chain. The other $N-1$ sites are, for a given
$\theta$, prepared in the ground state
$\ket{\text{gs}_\theta}_{2\ldots N}$ of $\hat H(\theta)$. The
initial state reads
$\ket{\psi_{\xi}(\theta,t=0)}=\ket{\xi}_1\otimes\ket{\text{gs}_\theta}_{2\ldots
N}$, see Fig.~\ref{fig:phases_scheme}(c). At $t=0$, we abruptly
switch on the interaction between the first and second spin and
let the system evolve, obtaining
$\ket{\psi_{\xi}(\theta,t)}=\text{exp}[-\text{i}t \hat
H(\theta)]\ket{\psi_\xi(\theta,0)}$ ($\hbar=1)$. At time $t$, the
quality of the transfer of $\ket{\xi}$ to the last spin of the
chain is evaluated by the fidelity
$\bra{\xi}\hat\rho_N(\theta,t)\ket{\xi}$ of retrieving $\ket{\xi}$ at the end of the chain. The reduced state of site $N$ is $\hat\rho_N(\theta,t)={\rm tr}_{1\ldots
N-1}\ket{\psi_\xi(\theta,t)}\bra{\psi_\xi(\theta,t)}$. The channel
fidelity \cite{horod:1999} is obtained from averaging over pure
states $\ket{\xi}$: \beq \label{F} F(\theta,t)=\int d\xi
\bra{\xi}\hat\rho_N(\theta,t)\ket{\xi}, \eeq where $d\xi$ is the
$\rm SU(3)$ invariant measure. For spin-$1$ always $1/3\leq
F\leq1$. We define $F(\theta)\equiv F(\theta,t^\star)$, where
$t^\star$ is the time for which the perturbation arrives at the
end of the chain for the first time (\ie, we ignore later maxima
from multiple reflections on the boundaries).

We also analyze
entanglement transfer, considering that the spin at the first
lattice site is entangled with a spin outside of the chain, say at
lattice site $0$ (see Fig.~\ref{fig:phases_scheme}(d)). As a
particular case we take a singlet state
$\ket{s}_{01}=(\ket{1,-1}_{01}-\ket{0,0}_{01}+\ket{-1,1}_{01})/\sqrt3$, such that initially
$\ket{\psi_s(\theta,0)}=\ket{s}_{01}\otimes\ket{\text{gs}_\theta}_{2\ldots
N}$. As before, at $t=0$ the coupling between sites $1$ and $2$ is
switched on, while site $0$ always remains uncoupled. We quantify
the transfer of entanglement to the end of the chain by the
logarithmic negativity \cite{plenio:2007} for sites $0$ and $N$
\beq \label{LN}\LN(\theta,t)=\text{log}_2||\hat
\rho^{\Gamma}_{0N}(\theta,t)||_1, \eeq 
where $\hat
\rho_{0N}(\theta,t)={\rm tr}_{1\ldots
N-1}\ket{\psi_s(\theta,t)}\bra{\psi_s(\theta,t)}$, $\Gamma$ denotes partial transposition, and $||A||_{1}=\sqrt{\textrm{tr}A^{\dagger}A}$.  As defined
here, $0\leq\LN\leq\log_2(3)$ gives an upper  bound for the number
of spin-$1/2$ singlets that can be distilled from $\hat
\rho_{0N}(\theta,t)$. Again we will use
$\LN(\theta)=\LN(\theta,t^{\star})$. 

\begin{figure}[t]
\begin{center}
\includegraphics[width=0.98\columnwidth]{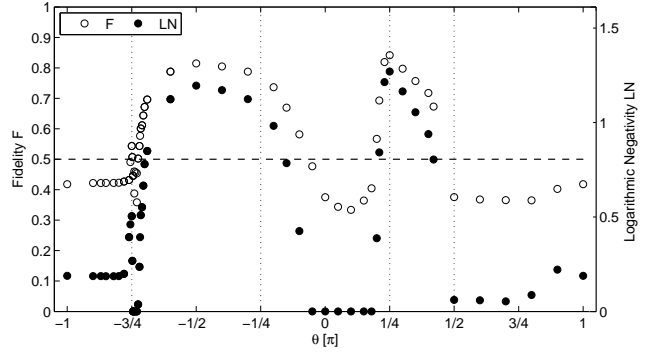}
\end{center}
\caption{Channel fidelity $F$ (open circles) and entanglement measured {\em via} logarithmic
negativity $\LN$ (filled circles) for the transfer between the
endpoints of a chain with $25$ sites (obtained with
MPS methods, using $D=20\ldots25$). The dashed horizontal line
denotes the maximal fidelity $F_{\rm class}=1/2$ which can be
obtained through classical communication. Vertical dashed lines indicate the quantum phase boundaries. (See
Fig.~\ref{fig:nematic} for a zoom into the region
$-0.76\pi\leq\theta\leq-0.7\pi$). } \label{fig:fidandln}
\end{figure}

We have computed $F(\theta)$ and $\LN(\theta)$ for chains of
up to $N=73$ sites using MPS simulations \cite{vidal:2004}. Our results
for $25$ sites and $-\pi\leq\theta<\pi$ are shown in
\figref{fig:fidandln}. The different quantum phases of the model
are well reflected in the transfer efficiency.
Before discussing several interesting points in detail, it is
useful to rewrite Eq. \eqref{Heisenberg} in terms of two-site
projectors $\hat P_{ij}^{(S_T)}=\sum_m\ket{S_T,m}_{ij}\bra{S_T,m}$
onto states with total spin $S_T$ ($m=S_T,...,-S_T$)
$$
\hat
H(\theta)=J\sum_{<ij>}H_{ij}(\theta)=J\sum_{<ij>} \lambda_{0}
\hat P^{(0)}_{ij}+\lambda_{1}
\hat P^{(1)}_{ij}+\lambda_{2}
\hat P^{(2)}_{ij}.
\label{eqn:hwithproj}
$$
Here $\lambda_0=-2\cos\theta+4\sin\theta,\,\lambda_1=-\cos\theta+\sin\theta,$
and $\lambda_2=\cos\theta+\sin\theta$.
We
start our discussion at the border of the critical phase and
follow the phase diagram in a counter-clockwise order.

\begin{figure}[t]
\begin{center}
\includegraphics*[width=0.99\columnwidth]{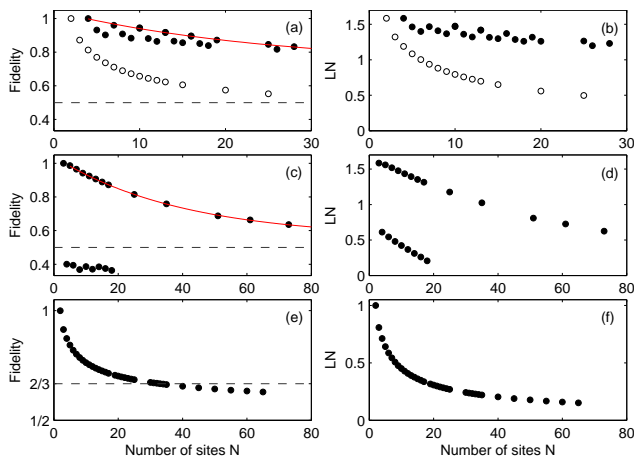}
\end{center}
\caption{Fidelity $F$ (left column) and logarithmic negativity $\LN$ (right column) versus chain
length $N$ for (a,b) the spin-$1$ chain at $\theta=\pi/4$; (c,d)
the spin-$1$ chain at $\theta=-\pi/2$; and (e,f) the spin-$1/2$
Heisenberg chain. Dark circles correspond to the channel prepared
in the ground state, open circles in (a,b) to a ferromagnetic
initial state. $F$ is fitted through an exponentially decaying
function. Dashed lines indicate the classical communication limit
$F_{\rm class}=1/2$ ($F_{\rm class}=2/3$) for spin-$1$ ($1/2$).
(Results obtained with MPS simulations using up to $D=25$ ($D=24$)
for $\theta=\pi/4$ ($-\pi/2$)).} \label{fig:trimer}
\end{figure}

{\it (a) Uimin-Lai-Sutherland point ($\theta=\pi/4$) and critical
phase ($\pi/4<\theta<\pi/2$)}.  Fidelity $F$ and logarithmic negativity $\LN$ attain their maximum
at $\theta=\pi/4$, and decrease towards the ferromagnetic phase. For $\theta=\pi/4$, the BBH takes a
particular simple form \beq
\hat{H}_{ij}(\pi/4)=\frac1{\sqrt2}\left[\vec S_i \vec S_j+(\vec
S_i\vec S_j)^2\right]
=\frac1{\sqrt2}\left[\hat{W}_{ij}+\frac{5}{7}\eins_{ij}\right],
\eeq where $\hat{W}_{ij}=(-1)^{2S}\sum_{S_T=0}^{2S}(-1)^{S_T}\hat
P_{ij}^{(S_T)}$ is the operator swapping sites $i$ and $j$ and $\eins_{ij}$ is the identity operator.
For $N=4$, the initial state is
$\ket{\psi_{\xi}(\pi/4,0)}=\ket{\xi}_1\otimes\ket{{\rm
gs}_{\pi/4}}_{234}$, with the trimer state 
\beq \ket{{\rm
gs}_{\pi/4}}_{234}=\ket{t}_{234}=\frac1{\sqrt6}\sum_P(-1)^{|P|}\mathbb{P}_P\ket{1,-1,0},
\eeq which has total spin zero
\cite{xian:1993} ($P$ runs over permutations of
$\left\{2,3,4\right\}$, $\mathbb{P}_P$ permutes the sites
accordingly). As
$\ket{\psi_{\xi}(\pi/4,Jt=\pi)}=\ket{t}_{123}\otimes\ket{\xi}_4$,
perfect transfer ($F=1$, $\LN=\log_2(3)$) occurs. For
$N>4$, transfer efficiency decreases (dark circles in
Fig.~\ref{fig:trimer} (a,b)). For small $N$ the transfer is better
for chains of length $N=1\mod3$, pointing to a trimerized order in
the ground state. As $N$ increases this difference vanishes. The average velocity
$v=N/t^{\star}$ at which the excitation propagates, grows as $N$
increases. From simulations for $N\leq28$, we extrapolate $\lim_{N\rightarrow\infty}v\approx1.59\,J\,{\rm
sites}$, a value close to the velocity of sound in the infinite system
$v_{\rm s}=2\pi/(3\sqrt2)\,J\,{\rm sites}\approx1.48\,J\,{\rm
sites}$ \cite{sutherland:1975}.

To emphasize the role of the initial state of the channel, we use
a ferromagnetic state $\ket{\psi_{\rm
F}(t=0)}=\ket{\xi}_1\otimes\ket{1,1,\ldots,1}_{2\ldots N}$ to
compare with. As $\hat{H}_{ij}$ swaps adjacent sites, this
reproduces the usual situation for spin-$1/2$ chains
\cite{bose:2003:b,subr:2004}. At time $t$ \beq \ket{\psi_{\rm
F}(\pi/4,t)}=\xi_{1}\ket{1\ldots1}+\sum_{n=1}^N\gamma_{1n}(t)\sum_{i=0,-1}\xi_i\ket{n_{i}},
\eeq where $\ket{n_{i}}$ represents the state with all spins in
state $0$, but the $n$th spin being in $i=0,-1$. The corresponding
probability amplitudes $\gamma_{1n}(t)$ can be calculated as an
infinite sum of Bessel functions \cite{childs:2002}. Already for
$N>2$, $|\gamma_{1N}(t^\star)|<1$, and thus also $F<1$. As visible
from the open circles in Fig.~\ref{fig:trimer} (a,b), the
transfer efficiency for the ferromagnetic initial state is much
below the efficiency of the chain initialized to its ground state.

{\it (b) Ferromagnetic phase
($\theta\in(\pi/2,\pi]\cup[-\pi,-3\pi/4)$)}. This region is
characterized by $\lambda_2<\lambda_{0},\lambda_1$. The
ground state has ferromagnetic order and broken rotational
symmetry: $\ket{{\rm
gs}_{\theta}(\vartheta,\varphi)}=\bigotimes_{i=2}^{N}\ket{1_{\vartheta,\varphi}}_i$. 
The state
$\ket{1_{\vartheta,\varphi}}_i$ has maximal spin projection in the direction specified by $(\vartheta,\varphi)$.
Throughout the whole phase transfer efficiency is very small. The point $\theta=\pi/2$ at the border to the critical
phase (where for the finite chain the ground state is
ferromagnetic) allows to identify two reasons: (i) for fixed
$(\vartheta,\varphi)$, $\ket{\xi}=\ket{0_{\vartheta,\varphi}}$
having vanishing $z$-projection in the corresponding direction is
not transported; (ii) the transfer of
$\ket{\xi}=\ket{-1_{\vartheta,\varphi}}$ is not \via swaps (as for spin-$1/2$ or at $\theta=\pi/4$),
but through an intermediate state.

{\it (c) Dimer phase ($-3\pi/4<\theta<-\pi/4$)}. Inside this
region, $F$ and $\LN$ increase strongly and reach a maximum at
$\theta=-\pi/2$. At this point $\lambda_0<\lambda_{1}=\lambda_2$,
thus the two-site ground state is a singlet. As
$\ket{\xi}_1\otimes\ket{s}_{23}\pm \ket{s}_{12}\otimes\ket{\xi}_3$
are both eigenstates of $\hat H(-\pi/2)$ (with different
eigenvalues), for $N=3$ perfect transfer occurs. For larger (odd)
$N$ the ground state of the last $N-1$ spins is dimerized
\cite{yip:2003,rizzi:2005}, \ie, the expectation values of
$\hat{s}_{i,i+1}=\ket{s}_{i,i+1}\bra{s}$ are different on even and
odd bonds. This is not the case for even $N$, and correspondingly
transfer fidelities vary strongly between even and odd $N$, see
Fig.~\ref{fig:trimer}(c,d).
Note that $\hat H(-\pi/2)=-\hat H(\pi/2)$, \ie, the ground state
of one model is the state with highest energy of the other, but
fidelities observed in both cases are very different. This
observation, which indeed can be made also at other points in the
phase diagram, is a clear manifestation of the dependence of the
transfer on the underlying magnetic order and the
nature of low excitations.

\begin{figure}[t]
\begin{center}
\includegraphics*[width=0.99\columnwidth]{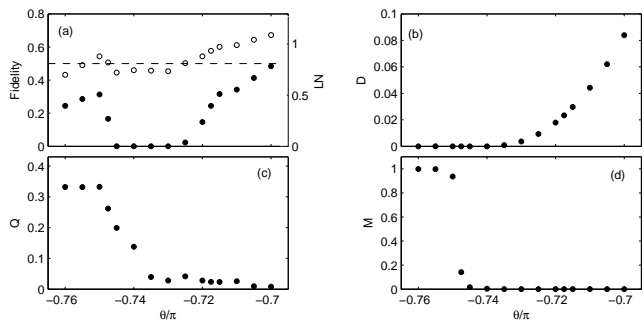}
\end{center}
\caption{ (a) $F$ (open circles) and $\LN$ (filled circles) around
the region for which a spin-nematic phase has been conjectured
(for $N=25$). The transfer efficiency is strongly reduced for
$-0.75\pi<\theta\lesssim0.72\pi$, as most clearly visible from
$\LN=0$. To characterize the ground state we plot (b) the
dimerization $D_{(N-1)/2}$ ($D_i=\ew{\hat H_{i,i+1}}-\ew{\hat
H_{i+1,i+2}}$), (c) the nematic order parameter
$Q=\max_{\Omega}\sum_{i=2}^N[\ew{(\vec{n}_{\Omega}\vec{S}_i)^2}-2/3]/(N-1)$
($\vec{n}_{\Omega}$ is the unit vector pointing in direction
$\Omega$), and (d) the magnetization
$M=\max_{\Omega}\sum_{i=2}^N\ew{\vec{n}_{\Omega}\vec{S}_i}/(N-1)$.
} \label{fig:nematic}
\end{figure}

{\it Conjectured spin-nematic phase
($-0.75\pi<\theta\lesssim-0.72\pi$)}. Near the border to the
ferromagnetic phase, Fig.~\ref{fig:fidandln} shows a remarkable
drop-off in transfer efficiency. Fig.~\ref{fig:nematic}(a) shows a
zoom into this region where a spin-nematic phase has been
suggested \cite{chubukov:1991}. This claim has been actively
discussed recently but a complete characterization of this region
is still under discussion
\cite{yip:2003,legeza:1997,ivanov:2003,buchta:2005,rizzi:2005,trebst:2006,porras:2006}. The dip for
$F$ and $\LN$ however is consistent with the results of
L{\"a}uchli {\it et al.} \cite{trebst:2006} and Porras {\it et
al.} \cite{porras:2006}, who found a qualitative change in the
excitation spectrum for values $-3\pi/4<\theta<\theta_C$, with
$\theta_C/\pi\approx-0.7\ldots-0.67$. To better characterize the
ground state in this region, we calculate dimerization, nematic
order parameter and magnetization (see
Fig.~\ref{fig:nematic}(b--d)).

{\it (d) Haldane phase ($-\pi/4<\theta<\pi/4$)}. A large
region in this phase shows poor transfer with fidelities close
to the lower limit $F=1/3$. At the AKLT point ($\tan\theta=1/3$) 
this can be understood from the equivalence of
the spin-$1$ chain to a model of two coupled spin-$1/2$ chains
\cite{legeza:1997}: while the couplings between the two chains are
completely symmetric, the AKLT ground state presents an asymmetry
due to the boundary spin-$1/2$s \cite{aklt:1987}. This may lead to
low transfer efficiency.

%

Let us finally discuss quantitative features by comparing to the
spin-$1/2$ chain without local engineering of couplings \footnote{
In spin-$1/2$ chains, locally engineering the couplings allows for
perfect state transfer \cite{christandl:2004}. In the spin-$1$ chain, arbitrarily perfect transfer can be realized
{\em via} adiabatic passage \cite{apdimer:2010}. }.
Contrary to Ref.~\cite{bose:2003:b}, where such a system was introduced, we always calculate $F$ and $\LN$ for
the {\it first} maximum, disregarding better values at
(possibly much) later times due to multiple
reflections and constructive interference %
\footnote{ In Ref.~\cite{bose:2003:b}, $F$ is maximized over times
up to $Jt=4000$. For $N=73$, the first maximum is obtained for
$Jt^\star\approx22.5$, \ie, more than $2$ orders of magnitude
earlier.
}. $F$ and $\LN$ for the spin-$1/2$ chain
initialized to a ferromagnetic state and using a Heisenberg
Hamiltonian are plotted in Fig.~\ref{fig:trimer} (e,f). A good
criterion to evaluate the efficiency of the channel is to compare
its fidelity with the highest fidelity for transmission of a
quantum state through a classical channel
$F_{\text{class}}=2/(d+1)$, being $d$ the dimension of the quantum
system \cite{horod:1999}. In particular, we can compute the
maximal length of the channel, \ie, the maximal number of sites
for which $F>F_{\text{class}}$. According to our numerical
results, for the spin-$1/2$ chain we obtain $F<F_{\rm class}=2/3$
for $N\leq33$. For spin-$1$ at $\theta=-\pi/2$ we find that
$F>F_{\rm class}=1/2$ for all our simulations, \ie, for $N\leq73$.
From fitting the fidelity through an exponentially decaying
function, we extrapolate
$F_{\infty}=\lim_{N\rightarrow\infty}F\approx0.56$, indicating
that the channel is {\it always} superior to any classical channel
(we cannot exclude a different decay for larger chains, though $N=73$ is well above the dimer
coherence length $N_{D}\approx20$ \cite{buchta:2005}). At
$\theta=\pi/4$, fidelities are even larger. Considering chains of
$N\leq28$ sites, we find that fidelity decays exponentially with
asymptotic limit $F_{\infty}\approx0.72$ well above the classical
value (again we cannot exclude a different decay for larger
chains).


Summarizing, we have studied state and entanglement transfer in
spin-$1$ Heisenberg chains. We have shown that the quality of transfer,
characterized by the average fidelity and the logarithmic
negativity, undergoes sharp changes at the phase boundaries. The
critical (gapless) and the dimerized (gapped) phase have high fidelities; from extrapolating our data the channel might even be non-classical for any number of sites.
In contrast, transfer efficiency is significantly lower in the ferromagnetic phase, 
and attains a minimum in the Haldane phase (around the  AKLT point)
and in a small region at the border 
between the dimerized and the ferromagnetic phase, 
where the existence of a spin-nematic order is controversially discussed.

{\it Acknowledgments} -- We thank J.J.~Garc{\'\i}a-Ripoll and
J.I.~Cirac for support in developing the MPS code, and M.A. Mart{\'\i}n-Delgado and R.~Mu\~noz-Tapia for discussions. We
acknowledge support from ESF PESC QUDEDIS and MEC under contracts EX2005-0830, AP2005-0595, FIS 2005-01369/014697, and SGR-00185.


\begin{thebibliography}{10}
%
%
%
%
%
%
%
%
%
%
%
%
%
%
%
%
%
%
%
%
%
%
%
%
%
%
%
%
%
%
%
%
%
%
%
%
%

\bibitem{osborne:2002}
T.J. Osborne and M.A. Nielsen,
\newblock Phys. Rev. A\ {\bf 66}, 032110\ (2002).

\bibitem{osterloh:2005}
A.~Osterloh {\em et al}.,
\newblock Nature\ {\bf 416}, 608\ (2005).

\bibitem{jaksch:1998}
D.~Jaksch {\em et al}.,\newblock Phys. Rev. Lett.\ {\bf 81}, 3108\
(1998); M.~Greiner {\em et al}., \newblock Nature\ {\bf 415}, 39\
(2002); D.~Porras and J.I. Cirac, \newblock Phys. Rev. Lett.\ {\bf
92}, 207901\ (2004).

\bibitem{demler:2002}
E.~Demler and F.~Zhou,
\newblock Phys. Rev. Lett.\ {\bf 88}, 163001\ (2002).

\bibitem{yip:2003}
S.K. Yip,
\newblock Phys. Rev. Lett.\ {\bf 90}, 250402\ (2003).

\bibitem{aklt:1987}
I.~Affleck {\em et al}.,
\newblock Phys. Rev. Lett.\ {\bf 59}, 799\ (1987).

\bibitem{uimin:1970}
G.V. Uimin,
\newblock JETP Lett.\ {\bf 12}, 225\ (1970);
C.K. Lai,
\newblock J. Math. Phys.\ {\bf 15}, 1675\ (1974).


\bibitem{sutherland:1975}
B.~Sutherland,
\newblock Phys. Rev. B\ {\bf 12}, 3795\ (1975).

\bibitem{tak:1982}
L.A. Takhtajan,
\newblock Phys. Lett. A\ {\bf 87}, 479\ (1982);
H.M. Babudjian,
\newblock Phys. Lett. A\ {\bf 90}, 479\ (1982).

%

\bibitem{chubukov:1991}
A.V. Chubukov,
\newblock Phys. Rev. B\ {\bf 43}, 3337\ (1991).

\bibitem{bose:2003:b}
S.~Bose,
\newblock Phys. Rev. Lett.\ {\bf 91}, 207901\ (2003).

\bibitem{subr:2004}
V.~Subrahmanyam,
\newblock Phys. Rev. A\ {\bf 69}, 034304\ (2004).

\bibitem{christandl:2004}
M.~Christandl {\em et al}.,
\newblock Phys. Rev. Lett.\ {\bf 92}, 187902\ (2004).

\bibitem{hartmann:2006}
M.J. Hartmann, M.E. Reuter, and M.B. Plenio,
\newblock New J. Phys.\ {\bf 8}, 94\ (2006).

\bibitem{li:2005}
Y.~Li {\em et al.},
\newblock Phys. Rev. A \ {\bf 71}, 022301\ (2005).

\bibitem{huo:2006}
M.X.~Huo,  {\em et al.},
\newblock quant-ph/060602.

\bibitem{horod:1999}
M.~Horodecki, P.~Horodecki, and R.~Horodecki,
\newblock Phys. Rev. A\ {\bf 60}, 1888\ (1999).

\bibitem{plenio:2007}
M.B. Plenio and S.~Virmani,
\newblock Quant. Inf. Comp.\ {\bf 7}, 1\ (2007).

\bibitem{vidal:2004}
G.~Vidal,
\newblock Phys. Rev. Lett.\ {\bf 93}, 040502\ (2004);
J.J. Garcia-Ripoll,
\newblock New J. Phys.\ {\bf 8}, 305\ (2006).

\bibitem{xian:1993}
Y.~Xian,
\newblock J. Phys. B: Condens. Matter\ {\bf 5}, 7489\ (1993).

\bibitem{childs:2002}
A.~Childs, E.~Farhi, and S.~Gutmann,
\newblock Quant. Inf. Proc.\ {\bf 1}, 35\ (2002);
A.J. Bessen,
\newblock quant-ph/0609128.

\bibitem{legeza:1997}
{\"O}.~Legeza, G.~F{\'a}th, and J.~S{\'o}lyom,
\newblock Phys. Rev. B\ {\bf 55}, 291\ (1997).

\bibitem{ivanov:2003}
B.A.~Ivanov and A.K.~Kolezhuk,
\newblock Phys. Rev. B\ {\bf 68}, 052401\ (2003).

\bibitem{rizzi:2005}
M.~Rizzi {\em et al}.,
\newblock Phys. Rev. Lett.\ {\bf 95}, 240404\ (2005).

\bibitem{buchta:2005}
K.~Buchta {\em et al}.,
\newblock Phys. Rev. B\ {\bf 72}, 054433\ (2005).

\bibitem{trebst:2006}
A.~L{\"a}uchli, G.~Schmid, and S.~Trebst,
\newblock cond-mat/0607173.

\bibitem{porras:2006}
D.~Porras, F.~Verstraete, and J.I. Cirac,
\newblock Phys. Rev. B\ {\bf 73}, 014410\ (2006).

\bibitem{apdimer:2010}
K.~Eckert, O.~Romero-Isart, and A.~Sanpera,
\newblock quant-ph/0702082.

\end{thebibliography}

\end{document}